\documentclass[pra,preprint,superscriptaddress,showpacs]{revtex4}

\usepackage{graphicx}
\usepackage{bm}                 
\usepackage{amsmath}            
\usepackage{amssymb}
\usepackage{verbatim}           
\usepackage{amsthm}             
\theoremstyle{plain}            

\def\bra#1{{\langle#1|}}
\def\ket#1{{|#1\rangle}}

\def\inner#1#2{{\langle#1|#2\rangle}}
\def\expect#1{{\langle#1\rangle}}
\def\e{{\rm e}}

\def\tr{{\rm Tr}}

\def\Ahat{{\hat A}}
\def\Adag{{\hat A}^\dagger}
\def\Ehat{{\hat E}}
\def\Edag{{\hat E}^\dagger}
\def\Shat{{\hat S}}

\def\U{{\hat U}}
\def\Udag{{\hat U}^\dagger}
\def\Zhat{{\hat Z}}

\def\Op{{\hat O}}
\def\id{{\hat I}}

\begin{document}

\title{Measuring polynomial functions of states}

\author{Todd A. Brun}
\email{tbrun@usc.edu}
\affiliation{Communication Sciences Institute, University of Southern California, Los Angeles, CA  90089-2565}

\date{September 2003}

\begin{abstract}
In this paper I show that any $m$th-degree polynomial function of the elements of the density matrix $\rho$ can be determined by finding the expectation value of an observable on $m$ copies of $\rho$, without performing state tomography.  Since a circuit exists which can approximate the measurement of any observable, in principle one can find a circuit which will estimate any such polynomial function by averaging over many runs.  I construct some simple examples and compare these results to existing procedures.
\end{abstract}


\maketitle

\section{Introduction}

Much of the practice of quantum information theory revolves around calculations of various {\it information measures}.  These are functions of the quantum state $\rho$ which quantify properties of interest, such as purity, entanglement, and distinguishability.  An important subset of these are {\it polynomial functions}.  Some polynomial quantities of potential interest are unitary invariants such as Kempe's invariant and the 3-tangle \cite{Kempe99,Coffman00,Carteret03a}, and measures of entanglement such as the $Q$ measure of Meyer and Wallach \cite{Meyer01,Brennen03}.  Some other quantities, which are not polynomials themselves, can either be approximated by polynomials (such as the quadratic approximation of the von~Neumann entropy) or are simple functions of such polynomials, such as the concurrence and the negativity of the partial transpose \cite{Wootters98,Peres96,Carteret03b}.  (I state what is meant by a polynomial function of $\rho$ more precisely below.)

A difficulty in estimating polynomial functions of the state is the linearity of quantum mechanics.  Given only one copy of a system in a particular quantum state $\rho$, any measurement can only depend linearly on $\rho$.  If it is possible to repeatedly prepare systems in the state $\rho$, we can overcome this obstacle by performing {\it quantum state tomography} \cite{Vogel89}:  by estimating the expectation values in the state $\rho$ of a set of observables, one can construct a classical description of the state; given that description, obviously any function of $\rho$ can be calculated.  However, many measurements are needed, of many different observables, since the number of parameters to be estimated grows like $d^2$ where $d$ is the dimension of the system's Hilbert space.

Recently there has been considerable interest in the problem of determining polynomial functions of quantum states (or simple functions of such polynomials); and a number of ingenious procedures have been put forward for measuring particular functions \cite{Sjoqvist00,Keyl01,Horodecki01a,Horodecki01b,Horodecki01c,Fiurasek01,Ekert02,Filip02,Alves03}.  These procedures try, by various clever techniques, to avoid estimating all the parameters needed for full state tomography; rather, they either estimate the polynomial function directly, or estimate a smaller set of parameters from which the polynomial can be calculated.  They do this by performing measurements on several copies of $\rho$ at once.

In this paper I demonstrate that for any $m$th-degree polynomial $f$ in the elements of the density matrix $\rho$ (on a Hilbert space $\cal H$), there are observables $\Op_f$ and $\Op_f'$ on ${\cal H}^{\otimes m}$ such that the expectation value of $\Op_f+i\Op_f'$ on $m$ copies of $\rho$ equals $f$:
\begin{equation}
f = \tr\{(\Op_f+i\Op_f') \rho^{\otimes m}\} \equiv \expect{\Op_f} + i \expect{\Op_f'}.
\end{equation}

\section{Polynomial functions}

What do I mean by a polynomial function on the elements of $\rho$?  Let us suppose that $\{\ket{i}\}$ is a basis for the Hilbert space $\cal H$, where $0 \le i < d$.  Then we can write $\rho$ in terms of that basis.
\begin{equation}
\rho = \sum_{i,j=0}^{d-1} \rho_{ij} \ket{i}\bra{j} .
\end{equation}
A polynomial function $f(\rho)$ of degree $m$ can be written
\begin{equation}
f(\rho) = \sum_{i_1,j_1,\ldots,i_m,j_m} c_{i_1 j_1 \ldots i_m j_m} \rho_{i_1j_1}\rho_{i_2j_2} \cdots \rho_{i_mj_m} ,
\label{polynomial}
\end{equation}
where the $\{c_{i_1 j_1 \ldots i_m j_m}\}$ are arbitrary complex constants.
There could also be terms of lower-order than $m$; but we can exploit the fact that $\tr\rho = \sum_i \rho_{ii} = 1$ to write them all as sums of terms of order $m$, so the form (\ref{polynomial}) is general.  This is easy to see:  consider a term of order $k<m$, proportional to $\rho_{i_1j_1}\cdots\rho_{i_kj_k}$.  We can rewrite this as a sum of terms of order $m$ by multiplying it by $m-k$ factors of $\tr\rho$:
\begin{equation}
\rho_{i_1j_1}\cdots\rho_{i_kj_k} = \sum_{i_{k+1},\ldots,i_m} \rho_{i_1j_1}\cdots\rho_{i_kj_k}\rho_{i_{k+1}i_{k+1}}\cdots\rho_{i_mi_m} .
\end{equation}

General polynomials of form (\ref{polynomial}) will not be invariant under a change of basis (though {\it some} polynomials of interest are invariant).  However, since a basis change is a linear transformation, such a transformation takes $m$th-degree polynomials to $m$th-degree polynomials.

Suppose we have $m$ copies of the system in state $\rho$; these copies have the joint state
\begin{equation}
\rho^{\otimes m} = \sum_{i_1,j_1,\ldots,i_m,j_m} \rho_{i_1j_1}\cdots\rho_{i_mj_m} \ket{i_1}\bra{j_1}\otimes \cdots \otimes \ket{i_m}\bra{j_m} .
\label{m_copy_state}
\end{equation}
Consider one term of the polynomial function $f(\rho)$ given by (\ref{polynomial}), proportional to $\rho_{i_1j_1}\cdots\rho_{i_mj_m}$.  We can write down an operator
\begin{equation}
\Ahat_{i_1j_1\cdots i_mj_m} = \ket{j_1}\bra{i_1}\otimes\cdots\otimes \ket{j_m}\bra{i_m} ,
\label{one_term_op}
\end{equation}
such that
\begin{equation}
\tr\{ \Ahat_{i_1j_1\cdots i_mj_m} \rho^{\otimes m} \} = \rho_{i_1j_1} \cdots \rho_{i_mj_m} ,
\label{one_term_expect}
\end{equation}
as we see by substituting (\ref{one_term_op}) and (\ref{m_copy_state}) in (\ref{one_term_expect}).
Since the full polynomial $f(\rho)$ is a linear combination of such terms, and the trace is a linear operation, we can find an operator $\Ahat_f$
\begin{equation}
\Ahat_f = \sum_{i_1,j_1,\ldots,i_d,j_d} c_{i_1 j_1 \ldots i_m j_m} \Ahat_{i_1j_1\cdots i_mj_m} ,
\label{polyop1}
\end{equation}
such that
\begin{equation}
f(\rho) = \expect{\Ahat_f} = \tr\{ \Ahat_f \rho^{\otimes m} \} .
\end{equation}

While this operator obviously exists, it will in general not be an observable---that will only be the case if $\Ahat_f$ is also Hermitian, $\Ahat_f = \Adag_f$.  However, we can trivially find a pair of operators which {\it are} observables,
\begin{eqnarray}
\Op_f &=& (\Ahat_f + \Adag_f)/2 , \nonumber\\
\Op'_f &=& -i(\Ahat_f - \Adag_f)/2 ,
\label{polyop2}
\end{eqnarray}
such that $\expect{\Ahat_f}=\expect{\Op_f} + i \expect{\Op'_f}$.

These operators will not be unique, in general.  If $\Ahat_f$ is an operator as defined above, then the operator obtained by permuting the order of the $m$ systems will also give the polynomial.  So will any linear combination of such permutations, so long as the coefficients of the linear combination sum to 1.  This freedom reflects the permutation symmetry of the state $\rho^{\otimes m}$

\section{Measuring polynomial functions}

If we have the ability to measure arbitrary observables, then we are done; we repeatedly prepare $\rho^{\otimes m}$ and measure $\Op_f$ and $\Op_f'$ until we have a good estimate of their expectation values.  If we do not have the ability to measure arbitrary observables, the fact that an observable exists whose expectation value yields the polynomial function is not very useful in itself.  However, even if we cannot measure arbitrary observables, we can design a circuit which will {\it simulate} such a measurement to any desired degree of accuracy.

The proof of this is trivial.  Let the observable $\Op_f$ have a basis of eigenstates $\ket{\phi_j}$ with corresponding eigenvalues $o_j$ (which are not necessarily distinct).  Let $\ket{j}$ be a {\it standard basis} which we are capable of measuring.  (For instance, in a system of q-bits this could be the computational basis.)  Then we can define a unitary transformation
\begin{equation}
\U_f = \sum_j \ket{j}\bra{\phi_j} ,
\end{equation}
which rotates the observable $\Op_f$ to be diagonal in the standard basis:
\begin{equation}
\Op_f = \sum_j o_j \ket{\phi_j}\bra{\phi_j} , \ \ \ 
\U_f \Op_f \Udag_f = \sum_j o_j \ket{j}\bra{j} .
\end{equation}
Clearly this new observable $\U_f\Op_f\Udag_f$ is straightforward to measure:  we just measure the state in the standard basis and give a value $o_j$ to outcome $j$. This gives us an expression for the expectation value of $\Op_f$,
\begin{eqnarray}
\tr\{\Op_f \rho^{\otimes m} \} &=&
  \tr\{(\U_f\Op_f\Udag_f) (\U_f\rho^{\otimes m}\Udag_f) \} \nonumber\\
&=& \sum_j o_j \bra{j}(\U_f\rho^{\otimes m}\Udag_f)\ket{j} .
\end{eqnarray}
All we need then is to be able to find a circuit which approximates the unitary transformation $\U_f$.  It is well-known that a circuit exists to approximate any unitary transformation to any desired degree of accuracy \cite{Divincenzo95}.  Our procedure now is as follows:  repeatedly prepare $m$ copies of the state $\rho$, and carry out the circuit which approximates $\U_f$.  Measure in the standard basis, and weight each outcome $j$ with the eigenvalue $o_j$.  Do the same for the observable $\Op_f'$.  We see, therefore, that in principle it is always possible to find a circuit which enables one to measure any polynomial function of $\rho$.

\section{Finding circuits}

This proof of principle is very far from being a proof that such a measurement is practical.  While such a circuit always exists, there is no guarantee that it will achieve the desired unitary transformation efficiently.  Indeed, most unitary transformations cannot be approximated efficiently \cite{Knill95}.  A protocol of this nature would only be worthwhile if it were substantially more efficient than performing quantum state tomography on $\rho$ and then calculating the function directly.  This requires comparing the cost of doing more measurements (in the case of state tomography) to the cost of preparing a more complicated state ($\rho^{\otimes m}$ instead of $\rho$) and performing some (possibly large) number of extra gates.

Even if it is possible, one must in most cases diagonalize a pair of complicated observables $\Op_f$ and $\Op'_f$ and find the corresponding eigenvalues and eigenvectors.  This may be a nontrivial task; if the system is of dimension $d$, then $\Op_f$ is a $d^m\times d^m$ Hermitian matrix.  However, for some applications this might be acceptable; the eigenvectors and eigenvalues could be calculated ``off-line,'' and then used to design a circuit which could be applied to many different states $\rho$.
There is no known efficient algorithm for finding the simplest circuit that produces a given unitary transformation.  Indeed, that is almost certain a computationally intractable problem in itself.  But in some cases, it may be possible to find a simple circuit to do the job.  Polynomials with a great deal of symmetry may simplify the circuits considerably.

Let us consider a simple example.  The first, and undoubtedly most immediately useful, is calculating the trace of $\rho^2$:
\begin{equation}
f = \tr\{\rho^2\} = \sum_{i,j} \rho_{ij}\rho_{ji} = \sum_{i,j} |\rho_{ij}|^2 .
\label{rho2}
\end{equation}
This quantity gives a useful measure of the {\it purity} of a state, with $\tr\{\rho^2\}=1$ for a pure state and $\tr\{\rho^2\}=1/d$ for the maximally mixed state; this is easier to calculate than the von~Neumann entropy.  For a bipartite pure state $\ket\Psi$, $\tr\{\rho_A^2\}$ gives a measure of the two subsystems' entanglement, where $\rho_A=\tr_B\{\ket\Psi\bra\Psi\}$ is the reduced density matrix for subsystem A.  From (\ref{rho2}) it is obvious that the correct observable is the swap operator:
\begin{equation}
\Op_f = \sum_{i,j} \ket{i}\bra{j} \otimes \ket{j}\bra{i} .
\end{equation}
This is a special case of the widely known fact that
\begin{equation}
\tr\{\rho^m\} = \tr\{ \Shat \rho^{\otimes m} \}
\end{equation}
where $\Shat$ is the cyclic shift operator:  $\Shat\ket{\psi_1}\otimes \cdots \otimes\ket{\psi_m}
= \ket{\psi_m} \otimes \ket{\psi_1} \otimes \cdots \otimes \ket {\psi_{m-1}}$.  In the case $m=2$, this is the pairwise swap, and is Hermitian as well as unitary, so that it is itself an observable.  It has eigenvalues $\pm 1$, as any operator which is both Hermitian and unitary must; and if our system is a single q-bit, a good eigenbasis is the usual Bell basis:
\begin{eqnarray}
\ket{\phi_0} &=& (\ket{00}+\ket{11})/\sqrt{2} , \ \ o_0 = 1 , \nonumber\\
\ket{\phi_1} &=& (\ket{00}-\ket{11})/\sqrt{2} , \ \ o_1 = 1 , \nonumber\\
\ket{\phi_2} &=& (\ket{01}+\ket{10})/\sqrt{2} , \ \ o_2 = 1 , \nonumber\\
\ket{\phi_3} &=& (\ket{01}-\ket{10})/\sqrt{2} , \ \ o_3 = -1 . 
\end{eqnarray}
Circuits for measuring the Bell states are widely known; Fig. 1 presents an example of such a circuit, consisting of a controlled-NOT gate, a Hadamard gate, and two single-bit meaurements in the computational basis.  In many cases it would be much easier to measure such an observable than to do full state tomography even on a single bit.

\begin{figure}[t]
\includegraphics{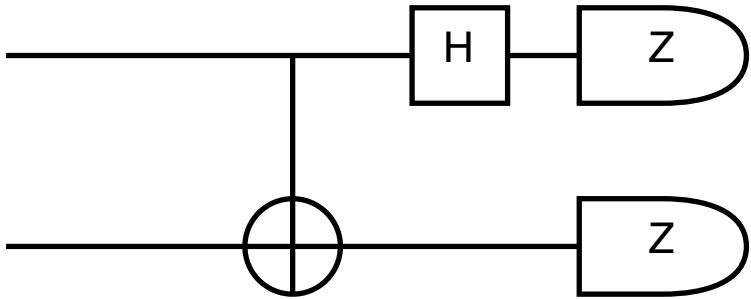}
\caption{\label{fig1}  Circuit for measuring two qubits in the Bell basis.}
\end{figure}

\section{Relationship to other approaches}

Other circuits have been presented in the literature for calculating other polynomial functions of the state \cite{Horodecki01b,Ekert02,Filip02,Alves03,Paz03}.  It is not difficult to see that these circuits give a method for estimating the expectation value of a suitable observable $\Op_f$, as described in this paper.

Let us look at this a bit more closely.  Consider the circuit in Fig. 2, which is a version of the circuit presented in \cite{Ekert02}.  In this circuit, a set of $m$ systems is prepared in the state $\rho^{\otimes m}$; an extra ``control bit'' is prepared in the state $\ket0$.  This control bit undergoes a Hadamard gate; it then serves as the control for a controlled-$\U$ gate (or circuit) with the $m$ systems, followed by another Hadamard.  Finally, the observable $\hat Z = \ket0\bra0 - \ket1\bra1$ is measured for the control bit.  It is straightforward to show that the expectation value of this measurement is $\expect{\hat Z}=(1/2)\tr\{(\U+\Udag)\rho^{\otimes m}\}$.  If $\U=\Shat$, for instance, then this would be equal to $\tr\{\rho^m\}$.  The operator $(1/2)(\U+\Udag)$ is Hermitian and therefore an observable.

Why does this work?  The values of the measurements can only be $1$ and $-1$; for most observables these will not be equal to {\it any} eigenvalue of $(1/2)(\U+\Udag)$.  In spite of this, the {\it expectation values} are exactly the same as if the observable $(1/2)(\U+\Udag)$ had been measured.  To see what is going on, let us suppose for the moment that $\U$ has eigenstates $\ket{\phi_j}$ with eigenvalues $\exp(i\theta_j)$.  We can always express the state of the $m$ systems in terms of these eigenstates,
\begin{equation}
\rho^{\otimes m} = \sum_{j,j'} R_{jj'}\ket{\phi_j}\bra{\phi_{j'}} .
\end{equation}
Suppose that we start the control bit and the $m$ systems in the joint state $\ket\Psi=\ket0\ket{\phi_j}$, and carry out the circuit in Fig. 2.  The resulting state will be
\begin{equation}
\ket{\Psi'} = \exp(i\theta_j/2)\left[ \cos(\theta_j/2)\ket0 - i \sin(\theta_j/2)\ket1 \right] \ket{\phi_j}
  \equiv \exp(i\theta_j/2)\ket{\gamma_j}\ket{\phi_j} .
\end{equation}
The expectation $\expect{\hat Z}$ in the control-bit state $\ket{\gamma_j}$ is $\cos^2(\theta_j/2)-\sin^2(\theta_j/2)=\cos(\theta_j)$ which is the $j$th eigenvalue of the operator $(1/2)(\U+\Udag)$.  A similar result is obtained for each eigenstate $\ket{\phi_j}$.  This circuit is therefore equivalent to the unitary transformation
\begin{equation}
\Ehat=\sum_j \e^{i\theta_j/2} \exp(-i\theta_j{\hat X}/2) \otimes \ket{\phi_j}\bra{\phi_j} .
\end{equation}
So a different way of understanding this circuit sees the control bit as a {\it target} bit:  if the $m$ systems are in the $j$th eigenstate of $\U$, the target bit is rotated about the $x$ axis from $\ket0$ to a new state $\ket{\gamma_j}$ whose $\hat Z$ expectation exactly matches the $j$th eigenvalue of the observable $(1/2)(\U+\Udag)$.  If we now start the $m$ systems in the state $\rho^{\otimes n}$, the different eigenstates of $\U$ will contribute with their appropriate weights $R_{jj}$:
\begin{eqnarray}
\tr\{ (\Zhat\otimes\id^{\otimes m})\Ehat(\ket0\bra0\otimes\rho^{\otimes m})\Edag \}
&=& \sum_{jj'} R_{jj'}\bra{\gamma_{j'}}\Zhat\ket{\gamma_j}\inner{\phi_{j'}}{\phi_j} \nonumber\\
&=& \sum_j R_{jj} \bra{\gamma_j}\Zhat\ket{\gamma_j} \nonumber\\
&=& \sum_j R_{jj} \bra{\phi_j}(1/2)(\U+\Udag)\ket{\phi_j} \nonumber\\
&=& \sum_{jj'} R_{jj'} \bra{\phi_{j'}}(1/2)(\U+\Udag)\ket{\phi_j} \nonumber\\
&=& \tr\{(1/2)(\U+\Udag)\rho^{\otimes m}\} .
\end{eqnarray}

Clearly, for any observable $\Op$ whose eigenvalues $o_j$ lie in the range $-1\le o_j \le 1$, we can find a unitary $\U$ such that $\Op = (1/2)(\U+\Udag)$.  For example, $\U=\Op+i\sqrt{\id-\Op^2}$ is such an operator.  If the eigenvalues of $\Op$ lie outside this range, we can construct a new observable $\Op/c$ where $c>0$ is larger than the absolute value of the largest eigenvalue of $\Op$.  We can then find a unitary $\U$ such that $(1/2)(\U+\Udag)=\Op/c$, and estimate the expectation value of $\Op/c$ by carrying out the circuit in Fig. 2.  This is an alternative and in some cases much more efficient way of estimating a polynomial function of $\rho$; in particular, it only requires measurements of a single bit.
Paz and Roncaglia \cite{Paz03} have constructed a programmable circuit of this type to estimate the expectation value of any observable.  This would obviously be sufficient to estimate any polynomial function of the state $\rho$.

\begin{figure}[t]
\includegraphics{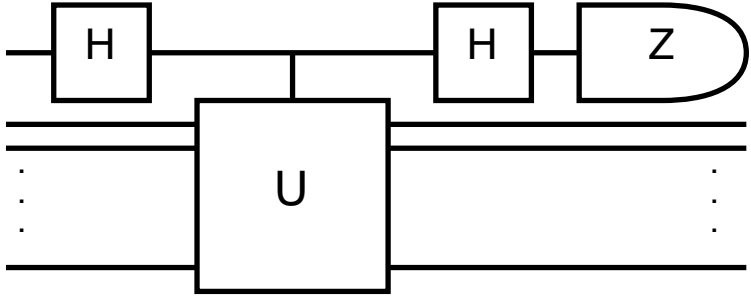}
\caption{\label{fig2}  Circuit for measuring $(1/2)\expect{\U+\Udag}$; after performing this circuit, the expectation value of $\hat Z$ for the first bit is equal to the expectation value of $(1/2)(\U+\Udag)$ in the initial state of the other $m$ systems.}
\end{figure}

\section{Conclusions}

The ability to measure arbitrary $m$-system observables and determine their expectation values allows the estimation of arbitrary $m$th-degree polynomials of the state, without having to perform quantum state tomography.  This is an enormous simplification, in principle; in some cases, it may give a considerable simplification in practice.  Whatever the practicalities, this equivalence is a useful fact, which may make possible improved protocols for estimating entanglement measures, polynomial approximations to the entropy, or other functions of the state.

\begin{acknowledgments}

I would like to acknowledge financial support from
the Martin A.~and Helen Chooljian Membership in Natural Sciences,
and DOE Grant No.~DE-FG02-90ER40542.  I appreciate helpful feedback on these ideas from Hilary Carteret and R\"udiger Schack.

\end{acknowledgments}


\end{document}